\documentclass[12pt]{article}

\usepackage{amsmath}
\usepackage{amsfonts}

\def\la{\langle}
\def\ra{\rangle}
\def\be{\begin{equation}}
\def\ee{\end{equation}}

\newtheorem{theorem}{Theorem}
\newtheorem{lemma}[theorem]{Lemma}

\newtheorem{definition}[theorem]{Definition}

\newtheorem{proposition}[theorem]{Proposition}

\begin{document}

\title{Frames of $p$--adic wavelets and orbits of the affine group}

\author{S.Albeverio, S.V.Kozyrev}

\maketitle

\begin{abstract}
The general construction of frames of $p$--adic wavelets is
described. We consider the orbit of a generic mean zero locally
constant function with compact support (mean zero test function)
with respect to the action of the $p$--adic affine group and show
that this orbit is a uniform tight frame. We discuss the relations
of this result with the multiresolution wavelet analysis.
\end{abstract}


\bigskip


\section{Introduction}

Wavelet analysis (see e.g. \cite{Daubechies}, \cite{Meyer},
\cite{Mallat}, \cite{NPS}) is the important method used in a wide
variety of applications from the theory of functions to signal
analysis.

A basis of wavelets in spaces of complex valued functions of a
$p$--adic argument was introduced and its relation to the spectral
analysis of $p$--adic pseudodifferential operators was investigated
in \cite{wavelets}. Other examples of $p$--adic wavelet bases were
considered in \cite{Benedetto}, \cite{Benedetto1}, \cite{ShelkSkop},
\cite{KhrShelkSkop}, \cite{AKhSh}. The relation between the
continuous and the discrete $p$--adic wavelet transform was
discussed in \cite{cont_wavelets}.

In the present paper we introduce the following general construction
of frames of $p$--adic wavelets. We act on an arbitrary function $f$
in $D_0(\mathbb{Q}_p)$ (mean zero locally constant functions with
compact support in the field $\mathbb{Q}_p$ of $p$--adic numbers) by
representations from the affine group (translations and dilations):
$$
G(a,b)f(x)=f^{a,b}(x)=|a|_p^{-{1\over 2}}f\left({x-b\over a}\right),
$$
where $a,b\in \mathbb{Q}_p$ and $a\ne 0$, $|a|_p$ being the
$p$--adic norm.

Due to local constancy of $f$ the orbit of $f$ will be a discrete
set of functions. We will show that, when the function $f$ is
generic (see the definition below), the orbit of $f\in
D_0(\mathbb{Q}_p)$ is a tight uniform frame, i.e. $\exists A>0$:
$\forall g\in L^{2}(\mathbb{Q}_p)$:
$$
\sum_{N}|\langle f^{(N)},g\rangle|^2=A\|g\|^2,
$$
where $f^{(N)}$ are elements of the orbit of $f$,
$\|f^{(N)}\|=\|f\|$, $\|\cdot\|$ is the $L^2$--norm.

In short the construction is as follows. Any function $f\in
D_0(\mathbb{Q}_p)$ has the form of a finite linear combination of
$p$--adic complex valued wavelets $\psi_{\gamma n j}$:
$$
f=\sum_{\gamma n j}C_{\gamma n j}\psi_{\gamma n j},\qquad  C_{\gamma
n j}\in \mathbb{C},\quad \psi_{\gamma n j}:\mathbb{Q}_p\to
\mathbb{C}.
$$
Then, we make several observations:

1) An orbit (with respect to the action of the affine group) of a
wavelet $\psi_{\gamma n j}$ is a set $\{e^{2\pi i
p^{-1}m}\psi_{\gamma' n' j'}\}$ of all products of wavelets and
roots of one of degree $p$;

2) The stabilizer of a wavelet $\psi_{\gamma n j}$ can be easily
computed;

3) We say that the function $f\in D_0(\mathbb{Q}_p)$ is generic, if
the stabilizer of action of the affine group on $f$ is an
intersection of stabilizers of wavelets which give contribution to
the above expansion of $f$. In this case the stabilizer of $f$ can
be easily computed;

4) Using the computed stabilizer of $f$, it is possible to construct
the set of representatives of the orbit $Gf$ (i.e. to parameterize
the orbit);

5) Using the constructed parametrization of the orbit $Gf$, one can
check that the orbit is a tight frame.

At the step 3 of our construction it is crucially important that due
to ultrametricity we can easily compute intersections, say
intersection of any finite number of $p$--adic balls is a ball (in
the real case an intersection of balls might have a quite
complicated structure).

We discuss the relation of the introduced construction of $p$--adic
frames and multiresolution analysis. We show that the
multiresolution construction in $p$--adic analysis corresponds to
the natural parametrization of the orbit of action of the affine
group on some mean zero test function, and that the multiresolution
construction possesses a family of generalizations (which correspond
to orbits of arbitrary mean zero test functions).

The results of the present paper support the point of view
\cite{cont_wavelets} that the main constructions of wavelet analysis
arise automatically as the properties of orbits of the $p$--adic
affine group and, moreover, the real wavelet analysis should be
considered as an analogue of these natural constructions of
$p$--adic analysis.

The structure of the present paper is as follows.

In Section 2 we describe the stabilizers and orbits with respect to
the action of the affine group of generic functions in
$D_0(\mathbb{Q}_p)$.

In Section 3 we prove that the orbit of a generic function $f\in
D_0(\mathbb{Q}_p)$ is a tight uniform frame.

In Section 4 we compare the structure of the constructed frame and
the definition of multiresolution analysis and discuss the relation
of muliresolution analysis with projective limits of $p$--adic
groups.

In Section 5 (the Appendix) we describe the necessary definitions
and notations.

\section{Orbits of the affine group}

If the function $f$ is a test function in $D(\mathbb{Q}_p)$, then
the orbit $G f$ of the function $f$ with respect to the action of
the affine group is a discrete set of test functions. In particular,
the following results take place.

\begin{lemma}\qquad {\sl

\noindent 1) The orbit $G\Omega(|\cdot|_p)$ is the set of all normed
characteristic functions of balls in $\mathbb{Q}_p$ and is in one to
one correspondence with the set of functions
$$
G(p^{-\gamma},p^{-\gamma}n)\Omega(|x|_p)=p^{-{\gamma\over
2}}\Omega(|p^{\gamma} x-n|_p),\qquad n\in \mathbb{Q}_p/\mathbb{Z}_p.
$$

\noindent 2) The stabilizer $G_{p^{-{\gamma\over
2}}\Omega(p^{\gamma}\cdot-n)}$ of $p^{-{\gamma\over
2}}\Omega(p^{\gamma}\cdot-n)$ contains all $g=(a,b)$ with $|a|_p=1$
and $b$ of the form
$$
b=p^{-\gamma}(n(1-a)+z),\qquad z\in \mathbb{Z}_p.
$$
In particular, the stabilizer of the unit ball $G_{\Omega}$ consists
of $b\in \mathbb{Z}_p$ and $a$ with $|a|_p=1$.

}

\end{lemma}

Here we assume that the elements of the factor group
$\mathbb{Q}_p/\mathbb{Z}_p$ are given by the representatives
(\ref{Qp/Zp}) in the corresponding equivalence classes.

\bigskip

\noindent{\it Proof}\qquad It is obvious that the affine group maps
a ball into a ball and is transitive on the balls. The first
statement of the lemma follows from the observation that the radii
of the balls in $\mathbb{Q}_p$ take values $p^{\gamma}$, $\gamma\in
\mathbb{Z}$, and the centers of the balls of the radius $p^{\gamma}$
can be chosen in the points $p^{-\gamma}n$, $n\in
\mathbb{Q}_p/\mathbb{Z}_p$.

Let us compute the stabilizer of $\Omega(|p^{\gamma}x-n|_p)$. For
$g=(a,b)\in G_{\Omega(|p^{\gamma}\cdot-n|_p)}$ we should have
$|a|_p=1$, because for $g$ in the stabilizer the action should not
change the diameter of the ball.

The point $p^{-\gamma}n$ belongs to the ball under consideration.
The image of this point with respect to the transformation from the
affine group is ${(p^{-\gamma}n-b)/ a}$. Therefore the necessary
(and sufficient) condition for $g=(a,b)$ to lie in the stabilizer is
$$
|{p^{-\gamma}n-b\over a}-p^{-\gamma}n|_p\le p^{\gamma},
$$
or
$$
|p^{-\gamma}n(1-a)-b|_p\le p^{\gamma}.
$$
Therefore the stabilizer $G_{\Omega(p^{\gamma}\cdot-n)}$ contains
all $g=(a,b)$ with $|a|_p=1$ and $b$ of the form
$$
b=p^{-\gamma}(n(1-a)+z),\qquad z\in \mathbb{Z}_p.
$$
In particular, the stabilizer of the unit ball $G_{\Omega}$ consists
of $b\in \mathbb{Z}_p$ and $a$ with $|a|_p=1$.

This finishes the proof of the lemma. $\square$

\bigskip

The next lemma describes the action of the affine group on $p$--adic
wavelets. The first statement of the lemma was proven in
\cite{cont_wavelets}.

\begin{lemma}\label{stab_of_wavelet}\qquad {\sl

\noindent 1) The orbit $G\psi(\cdot)$, where $\psi$ is the $p$--adic
wavelet:
$$
\psi(x)=\chi(p^{-1}x)\Omega(|x|_p),
$$
coincides with the set of products of functions from the wavelet
basis (\ref{basispaw}) and roots of 1 of degree $p$:
$$
G\psi=\{e^{2\pi ip^{-1}m}\psi_{\gamma n j}\},\qquad
m=0,1,\dots,p-1,\quad \gamma\in \mathbb{Z},\quad n\in
\mathbb{Q}_p/\mathbb{Z}_p,\quad j=1,\dots,p-1;
$$
where
$$
G(a,b)\psi=e^{2\pi ip^{-1}m}\psi_{\gamma n j},\qquad \gamma=\log_p
|a|_p,\quad j=(a|a|_p)^{-1}\,{\rm mod}\,p,
$$
$$
n=\{|a|_pb\},\quad m=\left(\left[(a|a|_p)^{-1}\,{\rm
mod}\,p\right](\{|a|_pb\}-|a|_pb)\right)  \,{\rm mod}\,p.
$$

The orbit is in one to one correspondence with the set of functions
$$
G(p^{-\gamma}j^{-1},p^{-\gamma}(n-mj^{-1}))\psi(x)=e^{2\pi
ip^{-1}m}\psi_{\gamma nj}(x).
$$

\noindent 2) The stabilizer $G_{\psi_{\gamma n j}}$ of the wavelet
$\psi_{\gamma n j}$ contains all $g=(a,b)$ with
$$
a=1\,{\rm mod}\,p,\qquad p^{\gamma}b=n(1-a) \,{\rm mod}\,p.
$$

}

\end{lemma}

\noindent{\it Proof}\qquad 1) Let us investigate the action of the
affine group on the wavelet $\psi$. We have
$$
\psi^{a,b}(x)=|a|_p^{-{1\over 2}}\psi\left({x-b\over
a}\right)=|a|_p^{-{1\over 2}}\chi\left(p^{-1}{x-b\over
a}\right)\Omega\left(\left| {x-b\over a} \right|_p\right).
$$
Setting $|a|_p=p^{\gamma}$ we get
$$
p^{-{\gamma\over 2}}\chi\left(p^{-1}{p^{\gamma}x-p^{\gamma}b\over
a|a|_p}\right)\Omega\left(|p^{\gamma}x-p^{\gamma}b|_p\right).
$$
Because for $a\ne 0$ we have $|a|a|_p|_p=1$, we can set
$\left(a|a|_p\right)^{-1}=j+z$, where $j=1,\dots p-1$ and $|z|_p<1$
(equivalently, $j=(a|a|_p)^{-1}\,{\rm mod}\,p$). Since the character
$\chi$ is locally constant and the argument of the character
satisfies $|p^{\gamma}x-p^{\gamma}b|\le 1$, we get
$$
p^{-{\gamma\over
2}}\chi\left(p^{-1}j(p^{\gamma}x-p^{\gamma}b)\right)\Omega\left(|p^{\gamma}x-p^{\gamma}b|_p\right).
$$
Let us set $p^{\gamma}b=n-j^{-1}m+\zeta$, $n\in
\mathbb{Q}_p/\mathbb{Z}_p$, $m=0,1,\dots,p-1$, $|\zeta|_p<1$ and
with $j$ as above. We get
$$
e^{2\pi ip^{-1}m}p^{-{\gamma\over
2}}\chi\left(p^{-1}j(p^{\gamma}x-n)\right)\Omega\left(|p^{\gamma}x-n|_p\right)=e^{2\pi
ip^{-1}m}\psi_{\gamma n j} (x),
$$
where
$$
\gamma=\log_p |a|_p,\quad j=(a|a|_p)^{-1}\,{\rm mod}\,p,\quad
n=\{|a|_pb\},\quad m=\left(\left[(a|a|_p)^{-1}\,{\rm
mod}\,p\right](\{|a|_pb\}-|a|_pb)\right)  \,{\rm mod}\,p.
$$

A direct computation gives
$$
G(p^{-\gamma}j^{-1},p^{-\gamma}(n-mj^{-1}))\psi(x)=p^{-{\gamma\over
2}}\psi\left(p^{\gamma}jx-jn+m\right)=
$$
$$
=p^{-{\gamma\over
2}}\chi(p^{-1}(p^{\gamma}jx-jn+m))\Omega(|p^{\gamma}jx-jn+m|_p)
=e^{2\pi ip^{-1}m}\psi_{\gamma nj}(x).
$$

2) Let us compute the stabilizer $G_{\psi_{\gamma n j}}$. We have
$$
G(a,b)\psi_{\gamma n j}(x)=p^{-{\gamma\over 2}} \chi\left(p^{-1}j
\left(p^{\gamma}{x-b\over a}-n\right)\right)
\Omega\left(\left|p^{\gamma} {x-b\over a}-n\right|_p\right).
$$
The diameter of the support of the wavelet is invariant iff
$|a|_p=1$. The index $j$ of the wavelet is invariant if we assume
$a|a|_p=1$. Therefore conservation of the indices $\gamma$ and $j$
implies for $a$ the condition
$$
a=1\,{\rm mod}\,p.
$$

Conservation of the support of the wavelet (of index $n$) implies
$$
p^{\gamma}b=n(1-a) \,{\rm mod}\,1.
$$
Conservation of the first multiplier in the wavelet (of the
character) gives the condition
$$
p^{\gamma}b=n(1-a) \,{\rm mod}\,p.
$$

This finishes the proof of the lemma. $\square$

\bigskip

Let the test function $f\in D_0(\mathbb{Q}_p)$ possess the following
(finite) expansion over $p$--adic wavelets:
\begin{equation}\label{test_function}
f=\sum_{\gamma n j}C_{\gamma n j}\psi_{\gamma n j},\qquad C_{\gamma
n j}\in \mathbb{C}.
\end{equation}

We say that the function $f$ is generic, if the stabilizer of action
of the affine group on $f$ is an intersection of stabilizers of
wavelets which give contribution to the above expansion of $f$.

\bigskip

\noindent{\bf Example}\qquad The example of a non--generic test
function can be constructed as follows. Consider the wavelet
$\psi_{-1, p^{-1}, j}$ (where $j$ can take values $1,\dots,p-1$)
with the support in the ball $x$: $|x-1|_p\le p^{-1}$. Consider the
transformation $G(a,b)$ from the affine group with $b=0$, $a$:
$|a|_p=1$, $|a-1|_p=1$, $a^{p-1}=1\,{\rm mod}\,p^2$ (for the proof
of the existence of $a$ satisfying the last condition see the
discussion below). Let us define
\begin{equation}\label{exotic_wavelet}
f(x)=\sum_{i=0}^{p-2}G^i(a,0)\psi_{-1, p^{-1},
j}(x)=\sum_{i=0}^{p-2}\psi\left(jp^{-1}\left({x\over
a^i}-1\right)\right)
\end{equation}
(i.e. $f$ is the sum of wavelets with supports in the maximal
subballs of the sphere $|x|_p=1$ given by iterations of the action
of $G(a,0)$ on $\psi_{-1, p^{-1}, j}$).

It is easy to see that $G(a,0)$ does not stabilize any of wavelets
in (\ref{exotic_wavelet}).

Let us prove that the function $f$ is invariant with respect to the
action of the described generator $G(a,0)$. Apply $G(a,0)$ to $f$.
We get
$$
G^i(a,0)f(x)=\sum_{i=1}^{p-1}\psi\left(jp^{-1}\left({x\over
a^i}-1\right)\right).
$$
We will have $G(a,0)f=f$ if
$$
\psi\left(jp^{-1}\left({x\over
a^{p-1}}-1\right)\right)=\psi\left(jp^{-1}\left({x}-1\right)\right).
$$

By the small Fermat theorem we have $a^{p-1}=1\,{\rm mod}\,p$.
Moreover, by the Hensel's lemma, since the derivative
$(a^{p-1}-1)'=(p-1)a^{p-2}\ne 0\,{\rm mod}\,p$ for $|a|_p=1$, there
exists $a$, $|a|_p=1$, such that $a^{p-1}=1\,{\rm mod}\,p^2$. Since
the function $\psi\left(jp^{-1}\left({x}-1\right)\right)$ is locally
constant with the diameter of local constancy equal to $p^{-2}$,
this proves that $f$ is invariant.

We see that this stabilization of the $f$ by action of the $G(a,0)$
is related with non--genericity of the $f$: if we multiply the
wavelets in (\ref{exotic_wavelet}) by generic coefficients, the
obtained function will not be stabilized by the action of $G(a,0)$.

\bigskip

The next lemma will give the parametrization of the action of the
affine group on an arbitrary generic test function in
$D_0(\mathbb{Q}_p)$.

\begin{lemma}\label{orbit}\qquad{\sl Assume that the function $f\in D_0(\mathbb{Q}_p)$
given by (\ref{test_function}) is generic. Then:

1) The stabilizer $G_f$ of the action of the affine group on $f$
given by (\ref{test_function}) consists of $g\in G$, $g=(a,b)$: $a$
belongs to the ball
\begin{equation}\label{stab_1}
|1-a|_p\le p^{-\gamma_A}={\rm min}\left[p^{-1},{{\rm
max}(p^{\gamma_i-1},p^{\gamma_j-1})\over
|p^{-\gamma_i}n_i-p^{-\gamma_j}n_j|_p}\right],
\end{equation}
where the minimum is taken over all $(\gamma_i,n_i)$ and
$(\gamma_j,n_j)$ in (\ref{test_function}) for which
$p^{-\gamma_i}n_i-p^{-\gamma_j}n_j\ne 0$, and $b$ satisfies
\begin{equation}\label{stab_2}
|b-p^{-\gamma_0}n_0(1-a)|_p\le p^{\gamma_0-1},
\end{equation}
where $\gamma_0$ is the minimal $\gamma$ in the expansion
(\ref{test_function}) and $n_0$ is the corresponding $n$.

2) The orbit $G f$ is in one to one correspondence with the set of
functions
\begin{equation}\label{orbita}
\{G(p^{\gamma}J,p^{\gamma}Jn)f\},\qquad \gamma\in \mathbb{Z},\quad
n\in \mathbb{Q}_p/p^{1-\gamma_0}\mathbb{Z}_p,
\end{equation}
\begin{equation}\label{J}
J=j+j_1p+j_2p^2+\dots+j_{\gamma_A-1}p^{\gamma_A-1},\quad
j=1,\dots,p-1,\quad j_i=0,\dots,p-1.
\end{equation}

}

\end{lemma}

Let us note that the indices $J$ form a group with respect to
multiplication $\,{\rm mod}\, p^{\gamma_A}$.

\bigskip

\noindent{\it Proof}\qquad 1) All the functions in the linear
combination (\ref{test_function}) are mutually orthogonal. Since the
action of the affine group is unitary, it conserves the
orthogonality.

Therefore the stabilizer of $f$ is the intersection of the
stabilizers of all the functions in the above linear combination. By
lemma \ref{stab_of_wavelet} we get that $g=(a,b)$ lies in the
stabilizer $G_f$ iff
\begin{equation}\label{stab}
a=1\,{\rm mod}\,p,\qquad p^{\gamma}b=n(1-a) \,{\rm mod}\,p
\end{equation}
for all pairs ($\gamma$, $n$) giving nonzero contributions
$C_{\gamma n j}$ to (\ref{test_function}).

Let us fix two pairs $(\gamma_1,n_1)$ and $(\gamma_2,n_2)$. We
choose $\gamma_1\le\gamma_2$. We describe the intersection of the
corresponding sets defined by (\ref{stab}), i.e. the solution of the
system
\begin{equation}\label{sys_2} a=1\,{\rm mod}\,p,\qquad
p^{\gamma_1}b=n_1(1-a) \,{\rm mod}\,p,\qquad p^{\gamma_2}b=n_2(1-a)
\,{\rm mod}\,p.
\end{equation}
For the fixed $a$ the sets of solutions of both equations for $b$
are the balls of the diameters $p^{\gamma_1-1}$, $p^{\gamma_2-1}$
and the centers $p^{-\gamma_1}n_1(1-a)$, $p^{-\gamma_2}n_2(1-a)$
correspondingly.

Due to the strong triangle inequality these balls have nonempty
intersection if
$$
|(p^{-\gamma_1}n_1-p^{-\gamma_2}n_2)(1-a)|_p\le {\rm
max}(p^{\gamma_1-1},p^{\gamma_2-1})=p^{\gamma_2-1}.
$$
If this condition is satisfied then $b$ belongs to the ball of the
diameter $p^{\gamma_1-1}$ with the center in
$p^{-\gamma_1}n_1(1-a)$.

We have proved that the solution of the system (\ref{sys_2})
consists of $a$ and $b$ satisfying
$$
|1-a|_p<1,\quad |1-a|_p\le {p^{\gamma_2-1}\over
|p^{-\gamma_1}n_1-p^{-\gamma_2}n_2|_p},\quad
|b-p^{-\gamma_1}n_1(1-a)|_p\le p^{\gamma_1-1},
$$
if $p^{-\gamma_1}n_1-p^{-\gamma_2}n_2\ne 0$, and
$$
|1-a|_p<1,\quad |b-p^{-\gamma_1}n_1(1-a)|_p\le p^{\gamma_1-1},
$$
if $p^{-\gamma_1}n_1-p^{-\gamma_2}n_2= 0$.

In general, consider the system of equations (\ref{stab}) for all
pairs ($\gamma$, $n$) giving nonzero contributions to
(\ref{test_function}). The solution of this system consists of $a$,
satisfying
$$
|1-a|_p\le {\rm min}\left[p^{-1},{{\rm
max}(p^{\gamma_i-1},p^{\gamma_j-1})\over
|p^{-\gamma_i}n_i-p^{-\gamma_j}n_j|_p}\right],
$$
where the minimum is taken over all $(\gamma_i,n_i)$ and
$(\gamma_j,n_j)$ for which $p^{-\gamma_i}n_i-p^{-\gamma_j}n_j\ne 0$,
and of $b$, satisfying
$$
|b-p^{-\gamma_0}n_0(1-a)|_p\le p^{\gamma_0-1},
$$
where $\gamma_0$ is the minimal $\gamma$ in the expansion
(\ref{test_function}) and $n_0$ is the corresponding $n$ (which, in
general, is not uniquely defined, but the corresponding set of $b$
is defined unambiguously).

2) Let us construct the parametrization of the orbit $Gf$ of the
function $f\in D_0(\mathbb{Q}_p)$ with respect to the action of the
affine group.

Consider the subgroup of dilations in the affine group
(transformations with $b=0$) and the system of representatives in
the orbit of $f$ with respect to transformations from this subgroup.
This system by (\ref{stab_1}) will be given by
\begin{equation}\label{para_orbit_0}
\{G(p^{\gamma}J,0)\},\quad \gamma\in Z,\quad
J=j+j_1p+j_2p^2+\dots+j_{\gamma_A-1}p^{\gamma_A-1},\quad
j=1,\dots,p-1,\quad j_i=0,\dots,p-1.
\end{equation}
Here $\gamma_A$ is given by (\ref{stab_1}).

Let us introduce the parametrization of the orbit of $f$ by
\begin{equation}\label{para_orbit}
\{G(p^{\gamma}J,p^{\gamma}Jn)f\},\qquad \gamma\in \mathbb{Z},\quad
n\in \mathbb{Q}_p/p^{1-\gamma_0}\mathbb{Z}_p
\end{equation}
and where $J$ is as in (\ref{para_orbit_0}).

To prove that the above formula indeed gives the parametrization of
the orbit $Gf$ of the affine group it is sufficient to show that any
transformation from the affine group possesses the unique expansion
$$
G(a,b)=G(a_1,b_1)G(a_0,b_0)=G(p^{\gamma}J,p^{\gamma}Jn)G(a_0,b_0),
$$
where $G(a_0,b_0)$ lies in the stabilizer of $f$ and
$G(a_1,b_1)=G(p^{\gamma}J,p^{\gamma}Jn)$ is the representative of
the orbit in formula (\ref{para_orbit}) (therefore
$G(a,b)f=G(a_1,b_1)f$).

To show this we define $\gamma$, $J$, $n$ in (\ref{para_orbit}) as
\begin{equation}\label{parametrization}
\gamma=\log_p |a|_p^{-1},\qquad J=a|a|_p \,{\rm
mod}\,p^{\gamma_A},\qquad n=\left(\left( ba_1^{-1} \right)\,{\rm
mod}\,p^{1-\gamma_0}\right),
\end{equation}
where $a_1=p^{\gamma}J$.

We have to prove that:

i) $G(a_0,b_0)=G^{-1}(a_1,b_1)G(a,b)$ lies in the stabilizer of $f$;

ii) for the different representatives of the orbit $G(a_1,b_1)$,
$G(a'_1,b'_1)$ the combination $$G^{-1}(a_1,b_1)G(a'_1,b'_1)$$ can
not belong to the stabilizer.

i) Since
$$
G(a_0,b_0)=G^{-1}(a_1,b_1)G(a,b)=G\left(a_1^{-1}a,(b-b_1)a_1^{-1}\right),
$$
we have
$$
a_0=a_1^{-1}a=a|a|_p\left(a|a|_p \,{\rm
mod}\,p^{\gamma_A}\right)^{-1}=1\,{\rm mod}\,p^{\gamma_A};
$$
$$
b_0=(b-b_1)a_1^{-1}=ba_1^{-1}-\left(\left( ba_1^{-1} \right)\,{\rm
mod}\,p^{1-\gamma_0}\right)=0\,{\rm mod}\,p^{1-\gamma_0}.
$$
Therefore $G(a_0,b_0)$ lies in the stabilizer.

ii) Since
$$
G^{-1}(a_1,b_1)G(a'_1,b'_1)=G\left(a_1^{-1}a'_1,(b'_1-b_1)a_1^{-1}\right),
$$
the above combination belongs to the stabilizer if
$$
a_1^{-1}a'_1=(p^{\gamma}J)^{-1}p^{\gamma'}J'=1\,{\rm
mod}\,p^{\gamma_A};
$$
$$
(b'_1-b_1)a_1^{-1}=n'a'_1a_1^{-1}-n=0\,{\rm mod}\,p^{1-\gamma_0};
$$
which is possible only if
$$
\gamma=\gamma',\quad J=J',\quad n=n'.
$$

This finishes the proof of the lemma. $\square$

\bigskip

\noindent{\bf Remark}\qquad The obtained in the above lemma
parametrization of the orbit $Gf$ is not unique. For instance, one
can, for each fixed $J$ and $\gamma$, choose the different
representatives $n$ in the equivalence classes in
$\mathbb{Q}_p/p^{1-\gamma_0}\mathbb{Z}_p$.

In particular, one can choose the parametrization of the elements of
the orbit $Gf$ in the form
\begin{equation}\label{para_orbit_1}
\{G(p^{\gamma}J,p^{\gamma}n)f\}
\end{equation}
where the indices are as in (\ref{para_orbit}).

\bigskip

\noindent{\bf Remark}\qquad For the above lemma it is crucially
important that in ultrametric spaces it is easy to compute
intersections. In particular, intersection of any finite number of
$p$--adic balls is a ball, while in the real case an intersection of
balls may have a quite complicated structure.

\section{Frames and orbits}

Consider the function $f\in D_0(\mathbb{Q}_p)$ defined by
(\ref{test_function}) and the set of representatives of action of
the affine group on this function defined by (\ref{orbita}):
$$
f^{(\gamma n J)}=G(p^{\gamma}J,p^{\gamma}Jn)f.
$$
We have the following lemma\footnote{ Here we use the standard
notations from quantum mechanics: $|\psi\ra$ denotes the element of
the Hilbert space $L^2(\mathbb{Q}_p)$ given by the function $\psi$,
$\la \psi|$ is the canonically conjugate linear bounded functional
on $L^2(\mathbb{Q}_p)$, and $|\psi\ra \la \phi|$,
$\|\psi\|=\|\phi\|=1$, is the operator in $L^2(\mathbb{Q}_p)$ which
maps $\phi$ onto $\psi$ and kills the orthogonal complement to
$\phi$. }:

\begin{lemma}\label{IWT}{\sl \qquad For $f\in D_0(\mathbb{Q}_p)$ given by (\ref{test_function})
the following series is proportional to the identity operator and is
equal to
\begin{equation}\label{inverse}
S=\sum_{\gamma n J}|f^{(\gamma n J)}\rangle \langle f^{(\gamma n
J)}|=\sum_{\gamma_1 n_1 j_1}|C_{\gamma_1 n_1
j_1}|^2p^{\gamma_A-\gamma_0+\gamma_1}
\end{equation}
(i.e. $S$ is an operator of multiplication by a number). The
summation over $\gamma$, $n$, $J$ runs over the parameters described
by (\ref{orbita}), (\ref{J}). The summation over $\gamma_1$, $n_1$,
$j_1$ runs over the indices of the wavelet basis. The $\gamma_A$ is
given by (\ref{stab_1}), $\gamma_0$ is the minimal $\gamma$ in
(\ref{test_function}).

The convergence of the series over $\gamma$, $n$, $J$ is understood
as the convergence of matrix elements in the wavelet basis (i.e. as
the weak convergence).}
\end{lemma}

\noindent{\it Proof}\qquad Since
$$
f=\sum_{\gamma n j}C_{\gamma n j}\psi_{\gamma n j},
$$
we have
$$
S=\sum_{\gamma n J}G(p^{\gamma}J,p^{\gamma}Jn) |f\rangle\langle
f|G^*(p^{\gamma}J,p^{\gamma}Jn)=
$$
\begin{equation}\label{S}
= \sum_{\gamma_1 n_1 j_1;\gamma_2 n_2 j_2}C_{\gamma_1 n_1
j_1}C^*_{\gamma_2 n_2 j_2}\sum_{\gamma n
J}G(p^{\gamma}J,p^{\gamma}Jn)|\psi_{\gamma_1 n_1 j_1}\rangle\langle
\psi_{\gamma_2 n_2 j_2}|G^*(p^{\gamma}J,p^{\gamma}Jn)
\end{equation}
We can change the order of the summation because the summation over
$(\gamma_1 n_1 j_1)$, $(\gamma_2 n_2 j_2)$ is finite and the series
over $(\gamma n J)$ applied to any wavelet gives a finite number of
non-vanishing terms.

Let us check that in the above sum over $(\gamma_1 n_1 j_1)$,
$(\gamma_2 n_2 j_2)$ only diagonal terms (i.e. terms with
$\gamma_1=\gamma_2$, $n_1=n_2$, $j_1=j_2$) give non--zero
contributions.

Consider the sum over $n$ (all the other indices are fixed):
$$
\sum_{n\in
\mathbb{Q}_p/p^{1-\gamma_0}\mathbb{Z}_p}G(p^{\gamma}J,p^{\gamma}Jn)|\psi_{\gamma_1
n_1 j_1}\rangle\langle \psi_{\gamma_2 n_2
j_2}|G^*(p^{\gamma}J,p^{\gamma}Jn)=
$$
$$
=G(p^{\gamma}J,0)\left[\sum_{n\in
\mathbb{Q}_p/p^{1-\gamma_0}\mathbb{Z}_p}G(1,n)|\psi_{\gamma_1 n_1
j_1}\rangle\langle \psi_{\gamma_2 n_2
j_2}|G^*(1,n)\right]G^*(p^{\gamma}J,0).
$$
The summation here runs over
$$
n=n_{\alpha}p^{\alpha}+n_{\alpha+1}p^{\alpha+1}+\dots+n_{-\gamma_0}p^{-\gamma_0},\qquad
n_j=0,\dots,p-1,\quad \alpha\le-\gamma_0.
$$
Note that by construction
$$
-\gamma_1\le-\gamma_0,\qquad -\gamma_2\le-\gamma_0,
$$
we also assume $-\gamma_1\le-\gamma_2$.

Therefore the summation over $n$ takes the form
$$
\sum_{n\in
\mathbb{Q}_p/p^{1-\gamma_0}\mathbb{Z}_p}G(1,n)|\psi_{\gamma_1 n_1
j_1}\rangle\langle \psi_{\gamma_2 n_2 j_2}|G^*(1,n)=
$$
$$
=\sum_{n\in
\mathbb{Q}_p/p^{-\gamma_2}\mathbb{Z}_p}\sum_{n_{-\gamma_2},n_{-\gamma_2}+1,\dots,n_{-\gamma_0}}G(1,n)|\psi_{\gamma_1
n_1 j_1}\rangle\langle \psi_{\gamma_2 n_2 j_2}|G^*(1,n)=
$$
$$
=p^{\gamma_0-\gamma_2}\sum_{n\in
\mathbb{Q}_p/p^{-\gamma_2}\mathbb{Z}_p}\sum_{n_{-\gamma_2}}G(1,n)|\psi_{\gamma_1
n_1 j_1}\rangle\langle \psi_{\gamma_2 n_2 j_2}|G^*(1,n).
$$
If $-\gamma_1<-\gamma_2$, then
$$
\sum_{n_{-\gamma_2}}G(1,n)|\psi_{\gamma_1 n_1 j_1}\rangle\langle
\psi_{\gamma_2 n_2 j_2}|G^*(1,n)=|\psi_{\gamma_1 n_1
j_1}\rangle\langle \psi_{\gamma_2 n_2
j_2}|\sum_{n_{-\gamma_2}=0,\dots,p-1}e^{2\pi
i{p^{-1}j_2n_{-\gamma_2}}}=0
$$
since the sum of roots of one of the degree $p$ is equal to zero.

If $-\gamma_1=-\gamma_2$ and $j_1\ne j_2$, we get
$$
\sum_{n_{-\gamma_1}}G(1,n)|\psi_{\gamma_1 n_1 j_1}\rangle\langle
\psi_{\gamma_1 n_2 j_2}|G^*(1,n)=|\psi_{\gamma_1 n_1
j_1}\rangle\langle \psi_{\gamma_1 n_2
j_2}|\sum_{n_{-\gamma_1}=0,\dots,p-1}e^{2\pi
i{p^{-1}(j_2-j_1)n_{-\gamma_1}}}=0.
$$
Therefore the only nonzero contributions in (\ref{S}) correspond to
contributions with $\gamma_1=\gamma_2$ and $j_1=j_2$.

Let us fix $\gamma_1=\gamma_2$, $j_1=j_2$ and $n_1\ne n_2$ and
consider the sum
$$
\sum_{J}G(J,0)|\psi_{\gamma_1 n_1 j_1}\rangle\langle \psi_{\gamma_1
n_2 j_1}|G^*(J,0).
$$
Note that at least one of $n_1$, $n_2$ is non zero.

We will investigate the sum over the subgroup of $J$ which leaves
invariant the supports of the both wavelets $\psi_{\gamma_1 n_1
j_1}$, $\psi_{\gamma_1 n_2 j_1}$ in the above expression. By the
previous lemma this sum contains $J$ which satisfy for non zero
$n_1$, $n_2$
$$
|J-1|_p\le\,{\rm
min}\,\left(|n_1|^{-1}_p,|n_2|^{-1}_p\right)=p^{-\gamma_B},
$$
and, for the case when one of $n_1$, $n_2$ is equal to zero, say
$n_1=0$,
$$
|J-1|_p\le|n_2|^{-1}_p=p^{-\gamma_B}.
$$
We get for such $J$
$$
G(J,0)\psi_{\gamma nj}(x)=p^{-{\gamma\over 2}} \chi(p^{-1}(j/J)
(p^{\gamma}x-nJ)) \Omega(|p^{\gamma} x-nJ|_p)=
$$
$$
=p^{-{\gamma\over 2}} \chi(p^{-1}j (p^{\gamma}x-nJ))
\Omega(|p^{\gamma} x-nJ|_p)=\psi_{\gamma nj}(x)e^{2\pi p^{-1}j
n(1-J)}.
$$

Let us compute the sum
$$
\sum_{J}G(J,0)|\psi_{\gamma_1 n_1 j_1}\rangle\langle \psi_{\gamma_1
n_2
j_1}|G^*(J,0)=\sum_{j,j_1,\dots,j_{\gamma_B}-1}\sum_{j_{\gamma_B}}\sum_{j_{\gamma_B+1},\dots,j_{\gamma_A-1}}
G(J,0)|\psi_{\gamma_1 n_1 j_1}\rangle\langle \psi_{\gamma_1 n_2
j_1}|G^*(J,0).
$$
We note that by (\ref{stab_1}) we have $\gamma_B\le \gamma_A-1$. The
sum over the indices $j_{\gamma_B}$,
$j_{\gamma_B+1},\dots,j_{\gamma_A-1}$ takes the form
$$
p^{\gamma_A-\gamma_B-1} |\psi_{\gamma_1 n_1 j_1}\rangle\langle
\psi_{\gamma_1 n_2 j_1}|\sum_{j_{\gamma_B}=0,\dots,p-1}e^{-2\pi
p^{-1}j_1 (n_1-n_2)j_{\gamma_B}}=0.
$$
Therefore nonzero contributions can only arise from the terms with
$n_1=n_2$.

We have proved that (\ref{S}) takes the form
$$
S= \sum_{\gamma_1 n_1 j_1}|C_{\gamma_1 n_1 j_1}|^2\sum_{\gamma n
J}G(p^{\gamma}J,p^{\gamma}Jn)|\psi_{\gamma_1 n_1 j_1}\rangle\langle
\psi_{\gamma_1 n_1 j_1}|G^*(p^{\gamma}J,p^{\gamma}Jn).
$$

Let us compute for the wavelet $\psi_{\gamma_1 n_1j_1}$ the number
of transformations
$$
G(p^{\gamma}J,p^{\gamma}Jn)\psi_{\gamma_1 n_1j_1}
$$
which act as a multiplication of $\psi_{\gamma_1 n_1j_1}$ by a root
of one (and therefore leave the rank one projection operator
$|\psi_{\gamma_1 n_1j_1}\rangle\langle\psi_{\gamma_1 n_1j_1}|$
unchanged). To obtain a multiplication of the wavelet by a constant
we should take $\gamma=0$ and $j=1$ in the formula (\ref{J}) for $J$
(i.e. $J=1\,{\rm mod}\,p$). We get the transformation
$$
G(J,Jn)\psi_{\gamma_1 n_1j_1}(x) =p^{-{\gamma_1\over 2}}
\chi\left(p^{-1}{j_1\over J} \left(p^{\gamma_1}
x-J(p^{\gamma_1}n+n_1)\right)\right) \Omega\left(\left|p^{\gamma_1}
x-J(p^{\gamma_1}n+n_1)\right|_p\right)=
$$
$$
=p^{-{\gamma_1\over 2}} \chi\left(p^{-1}j_1 \left(p^{\gamma_1}
x-J(p^{\gamma_1}n+n_1)\right)\right) \Omega\left(\left|p^{\gamma_1}
x-J(p^{\gamma_1}n+n_1)\right|_p\right).
$$

In order to get the support of the wavelet unchanged we need
$$
Jp^{\gamma_1}n=(1-J)n_1\,{\rm mod}\, 1
$$
or equivalently
\begin{equation}\label{degeneracy}
|Jp^{\gamma_1}n-(1-J)n_1|_p\le 1.
\end{equation}

The summation runs over
$$
n=n_{\alpha}p^{\alpha}+n_{\alpha+1}p^{\alpha+1}+\dots+n_{-\gamma_0}p^{-\gamma_0},\qquad
n_j=0,\dots,p-1,\quad \alpha\le-\gamma_0;
$$
$$
J=1+j_1p+j_2p^2+\dots+j_{\gamma_A-1}p^{\gamma_A-1},\quad
j_i=0,\dots,p-1.
$$

Let us compute the number of $n$ satisfying (\ref{degeneracy}) for
fixed $J$. Since there always exists the solution of
(\ref{degeneracy}) given by
$n_0=\left(J^{-1}p^{-\gamma_1}(1-J)n_1\right)\,{\rm
mod}\,p^{1-\gamma_0}$, the set of solutions will be given by $n$:
$|n-n_0|_p\le p^{\gamma_1}$. There are $p^{-\gamma_0+\gamma_1+1}$
such $n$.

Since there are $p^{\gamma_A-1}$ possible meanings of $J$, in total
we have $p^{\gamma_A-\gamma_0+\gamma_1}$ solutions of
(\ref{degeneracy}).

Since the set of transformations (\ref{orbita}) is transitive on
vectors from the basis of wavelets, the following series of
transformations of the projection operator by
$G(p^{\gamma}J,p^{\gamma}Jn)$ is proportional to the identity
operator:
$$
\sum_{\gamma n J}G(p^{\gamma}J,p^{\gamma}Jn)|\psi_{\gamma_1 n_1
j_1}\rangle\langle \psi_{\gamma_1 n_1
j_1}|G^*(p^{\gamma}J,p^{\gamma}Jn)=p^{\gamma_A-\gamma_0+\gamma_1}.
$$

This implies for $S$ the expression
$$
S= \sum_{\gamma_1 n_1 j_1}|C_{\gamma_1 n_1 j_1}|^2\sum_{\gamma n
J}G(p^{\gamma}J,p^{\gamma}Jn)|\psi_{\gamma_1 n_1 j_1}\rangle\langle
\psi_{\gamma_1 n_1 j_1}|G^*(p^{\gamma}J,p^{\gamma}Jn)=\sum_{\gamma_1
n_1 j_1}|C_{\gamma_1 n_1 j_1}|^2p^{\gamma_A-\gamma_0+\gamma_1}.
$$

This finishes the proof of the lemma. $\square$

\bigskip

The next theorem describes frames of $p$--adic wavelets given by
orbits of generic functions from $D_0(\mathbb{Q}_p)$ with respect to
the action of the affine group.

\begin{theorem}\label{main_theorem}\qquad {\sl Let $f\in D_0(\mathbb{Q}_p)$ is generic and let $f$ possess the
expansion over wavelets
\begin{equation}\label{exp}
f=\sum_{\gamma n j}C_{\gamma n j}\psi_{\gamma n j};\qquad \gamma\in
\mathbb{Z},\quad n\in \mathbb{Q}_p/\mathbb{Z}_p,\quad j=1,\dots,p-1.
\end{equation}

Then the orbit of $f$ with respect to action of the affine group
possess the following properties:

1) the orbit is given by the set of functions:
\begin{equation}\label{orbit_1}
f^{(\gamma n J)}=G(p^{\gamma}J,p^{\gamma}Jn)f,\qquad \gamma\in
\mathbb{Z},\quad n\in \mathbb{Q}_p/p^{1-\gamma_0}\mathbb{Z}_p,
\end{equation}
$\gamma_0$ is the minimal $\gamma$ in (\ref{exp}),
$$
J=j+j_1p+j_2p^2+\dots+j_{\gamma_A-1}p^{\gamma_A-1},\quad
j=1,\dots,p-1,\quad j_i=0,\dots,p-1;
$$
$$
p^{-\gamma_A}={\rm min}\left[p^{-1},{{\rm
max}(p^{\gamma_i-1},p^{\gamma_j-1})\over
|p^{-\gamma_i}n_i-p^{-\gamma_j}n_j|_p}\right],
$$
where the minimum is taken over all $(\gamma_i,n_i)$ and
$(\gamma_j,n_j)$ in (\ref{exp}) for which
$p^{-\gamma_i}n_i-p^{-\gamma_j}n_j\ne 0$;

2) the orbit is a uniform and tight frame in $L^2(\mathbb{Q}_p)$:
the norms of all $f^{(\gamma n J)}$ are equal and for any $g\in
L^2(\mathbb{Q}_p)$
\begin{equation}\label{frame}
\sum_{\gamma n J} |\langle g,f^{(\gamma n
J)}\rangle|^2=\|g\|^2\sum_{\gamma n j}|C_{\gamma n
j}|^2p^{\gamma_A-\gamma_0+\gamma}.
\end{equation}
}
\end{theorem}

Note that on the LHS of (\ref{frame}) the summation runs over the
elements of the orbit of $f$ and on the RHS the summation runs over
the elements of the wavelet basis.

\bigskip

\noindent{\it Proof}\qquad The first statement of the theorem is
proven in lemma \ref{orbit}.

The second statement follows from lemma \ref{IWT}: for $g\in
D_0(\mathbb{Q}_p)$ one can see that by lemma \ref{IWT}
$$
\sum_{\gamma n J} |\langle g,f^{(\gamma n J)}\rangle|^2=\langle
g,Sg\rangle=S\|g\|^2=\|g\|^2\sum_{\gamma n j}|C_{\gamma n
j}|^2p^{\gamma_A-\gamma_0+\gamma}.
$$
Since $D_0(\mathbb{Q}_p)$ is dense in $L^2(\mathbb{Q}_p)$ this
finishes the proof of the theorem. $\square$

\bigskip

\noindent {\bf Example 1}\qquad Consider the case
$f(x)=\psi(x)=\chi(p^{-1}x)\Omega(|x|_p)$. In this case by lemma
\ref{stab_of_wavelet} the orbit of $f$ is given by the set
$\{e^{2\pi i p^{-1}m}\psi_{\gamma n j}\}$ of products of wavelets
from the basis (\ref{basispaw}) of \cite{wavelets} and roots of one
of the degree $p$. It is easy to see that the bound of the frame
will be equal to $p$:
$$
\sum_{\gamma n J} |\langle g,f^{(\gamma n J)}\rangle|^2=p\|g\|^2.
$$

\bigskip

\noindent {\bf Example 2}\qquad Consider the more general example
when
\begin{equation}\label{fixed_scale}
f=\sum_{n j}C_{n j}\psi_{\gamma n j}
\end{equation}
i.e. $f$ is equal to the linear combination of wavelets with the
fixed scale $\gamma$.

In this case by (\ref{frame}) one gets for the bound of the
corresponding frame $\{f^{(\gamma n J)}\}$ the expression
$$
p^{\gamma_A}\sum_{n j}|C_{n j}|^2=p^{\gamma_A}\|f\|^2
$$
where $\gamma_A\ge 1$ (the minimal $\gamma_A= 1$ was considered in
the previous example when $f=\psi$).

Note that all examples of $p$--adic wavelet bases built in
\cite{ShelkSkop}, \cite{KhrShelkSkop} correspond to the case
(\ref{fixed_scale}).

\section{Relation to the multiresolution analysis}

The frame $\{G(p^{\gamma}J,p^{\gamma}Jn)f\}$ described in theorem
\ref{main_theorem}, is constructed by translations (which correspond
to the index $n$) and dilations (described by the indices $\gamma$
and $J$) of the generic function $f\in D_0(\mathbb{Q}_p)$. Therefore
it is natural to discuss the set $\{G(p^{\gamma}J,p^{\gamma}Jn)f\}$
as a frame of wavelets.

By definition,  a frame of $p$--adic wavelets is given by the set of
translations and dilations
\begin{equation}\label{paw}
G(p^{\gamma_1},p^{\gamma_1}n_1)f^{(J_1)},\qquad \gamma_1\in
\mathbb{Z}, \quad n_1\in \mathbb{Q}_p/\mathbb{Z}_p,
\end{equation}
and $\{f^{(J_1)}\}$ is a finite set of mean zero functions in
$L^2(\mathbb{Q}_p)$ (in general, we do not assume that the functions
in $\{f^{(J_1)}\}$ are shifts or dilations of some fixed function).

\begin{proposition}\qquad
{\sl Either the orbit $Gf$, $f\in D_0(\mathbb{Q}_p)$ described in
theorem \ref{main_theorem} satisfies itself the definition
(\ref{paw}) of the frame of wavelets, or some finite number of
copies of the orbit $Gf$ satisfies definition (\ref{paw}) of the
frame of wavelets. }
\end{proposition}

\noindent{\it Proof}\qquad Let us choose for $Gf$ the
parametrization (\ref{para_orbit_1}):
$Gf=\{G(p^{\gamma}J,p^{\gamma}n)f\}$.

By the remark after the lemma \ref{orbit} the orbit $Gf$ with
respect to the affine group of the function $f\in D_0(\mathbb{Q}_p)$
is a tight frame $\{G(p^{\gamma}J,p^{\gamma}n)f\}$ which possesses
the parametrization (\ref{para_orbit_1}). Here the scale $\gamma$
takes integer values as in (\ref{paw}), the translations  $n\in
\mathbb{Q}_p/p^{1-\gamma_0}\mathbb{Z}_p$ take an infinite number of
possible values and the index $J$ enumerates some finite set of
dilations by $p$--adic numbers with the norm one.

The translations $n$ run in
$\mathbb{Q}_p/p^{1-\gamma_0}\mathbb{Z}_p$, i.e. depending on $n_0$
either $\mathbb{Q}_p/p^{1-\gamma_0}\mathbb{Z}_p$ is a factor group
of $\mathbb{Q}_p/\mathbb{Z}_p$, or $\mathbb{Q}_p/\mathbb{Z}_p$ is a
factor group of $\mathbb{Q}_p/p^{1-\gamma_0}\mathbb{Z}_p$, in both
cases the index of the factor group (number of elements in an
equivalence class) will be finite.

Consider the following cases:

1) Let  $\mathbb{Q}_p/\mathbb{Z}_p$ be a factor group of
$\mathbb{Q}_p/p^{1-\gamma_0}\mathbb{Z}_p$.

Let us prove that the frame $Gf$ possesses parametrization
(\ref{paw}) for some finite set $\{f^{(J_1)}\}$ of functions in
$D_0(\mathbb{Q}_p)$.

Let us rewrite the parametrization (\ref{para_orbit_1}) of the orbit
$Gf$ in the form
$$
G(p^{\gamma_1},p^{\gamma_1}n_1)f^{(J_1)}(x)=G(p^{\gamma_1},p^{\gamma_1}n_1)f\left({x-m\over
J}\right)=G(p^{\gamma_1}J,p^{\gamma_1}(n_1+m))f(x).
$$
Here $\gamma_1\in \mathbb{Z}$, $n_1\in \mathbb{Q}_p/\mathbb{Z}_p$,
$m\in \mathbb{Z}_p/p^{1-\gamma_0}\mathbb{Z}_p$, $J$ is defined by
(\ref{J}).

This means that $J_1$ takes values $(J,m)$ and
$$
f^{(J_1)}(x)=f\left({x-m\over J}\right).
$$

Since the set of possible $J_1=(J,m)$ is finite, this proves that
$Gf$ satisfies definition (\ref{paw}).

2) Let $\mathbb{Q}_p/p^{1-\gamma_0}\mathbb{Z}_p$ be a factor group
of $\mathbb{Q}_p/\mathbb{Z}_p$. In this case
\begin{equation}\label{frame_1}
G(p^{\gamma_1},p^{\gamma_1}n_1)f^{(J_1)}(x)=G(p^{\gamma_1},p^{\gamma_1}n_1)f\left({x\over
J_1}\right),
\end{equation}
$\gamma_1\in \mathbb{Z}$, $n_1\in \mathbb{Q}_p/\mathbb{Z}_p$, $J_1$
is defined by (\ref{J}).

Since $f$ is invariant under the action of translations from
$p^{1-\gamma_0}\mathbb{Z}_p/\mathbb{Z}_p$, the above set of
functions coincides with the orbit $Gf$ taken $p^{\gamma_0-1}$
times. Therefore (\ref{frame_1}) is a tight frame.

This finishes the proof of the proposition. $\square$

\bigskip

The above proposition is related to the fact that for any generic
$f\in D_0(\mathbb{Q}_p)$ the stabilizer of the action of the affine
group has a similar form and therefore the orbits of any $f\in
D_0(\mathbb{Q}_p)$ possess a simple parametrization. Let us stress
that this observation is related to the strong triangle inequality
which is the crucial property of $p$--adic analysis.

Let us discuss the relation of the constructed frames to $p$--adic
multiresolution analysis. The following $p$--adic analogue of the
multiresolution approximation was discussed in \cite{cont_wavelets},
\cite{ShelkSkop}.

\begin{definition}\label{padicMRA}\qquad {\sl A multiresolution approximation of
$L^2(\mathbb{Q}_p)$ is a decreasing sequence $V_\gamma$, $\gamma\in
\mathbb{Z}$, of closed linear subspaces of $L^2(\mathbb{Q}_p)$ with
the following properties:

1)
$$
\bigcap_{-\infty}^{+\infty}V_{\gamma}=\{0\},\qquad
\bigcup_{-\infty}^{+\infty}V_{\gamma}\quad {\rm is~dense~in}\quad
L^2(\mathbb{Q}_p);
$$
2)  for all $f\in L^2(\mathbb{Q}_p)$ and all $\gamma\in \mathbb{Z}$,
$$
f(\cdot)\in V_{\gamma}  \Longleftrightarrow  f(p^{-1}\cdot)\in
V_{\gamma-1};
$$
3) for all $f\in L^2(\mathbb{Q}_p)$ and all $n\in
\mathbb{Q}_p/\mathbb{Z}_p$,
$$
f(\cdot)\in V_0 \Longleftrightarrow f(\cdot-n)\in V_0;
$$
4)  there exists a function $\phi\in V_0$ (called the scaling
function) such that the sequence
$$
\phi(\cdot-n),\quad n\in \mathbb{Q}_p/\mathbb{Z}_p
$$
is an orthonormal basis of the space $V_0$.

}

\end{definition}

The orthogonal complement in the space $V_{\gamma-1}$ of the
subspace $V_{\gamma}$ is called the wavelet space $W_{\gamma}$. The
space $L^2(\mathbb{Q}_p)$ is the orthogonal sum of $W_{\gamma}$,
$\gamma\in Z$. The spaces $W_{\gamma}$ satisfy the property
analogous to the property 2 of the above definition:
\begin{equation}\label{prop_2}
f(\cdot)\in W_{\gamma}  \Longleftrightarrow  f(p^{-1}\cdot)\in
W_{\gamma-1};
\end{equation}

\bigskip

\noindent {\bf Example}\qquad An example of the multiresolution
construction is given by the scaling function
$\phi(x)=\Omega(|x|_p)$. In this case $V_{\gamma}$ contains test
functions with the diameter of local constancy $p^{\gamma}$. The
wavelet space $W_{\gamma}$ possesses the basis $\{\psi_{\gamma n
j}\}$ of $p$--adic wavelets with the fixed index $\gamma$ and $n\in
\mathbb{Q}_p/\mathbb{Z}_p$, $j=1,\dots,p-1$.

\bigskip

Let us compare the above definition with the structure of the orbit
of generic $f\in D_0(\mathbb{Q}_p)$ with respect to action of the
affine group.

In the case of the $p$--adic wavelet $f=\psi$ we get the orbit
$\{e^{2\pi i p^{-1}m}\psi_{\gamma n j}\}$. Roughly speaking this
orbit is given by $p$ copies of the basis of $p$--adic wavelets.
Defining $W_{\gamma}$ as the linear span of vectors $e^{2\pi i
p^{-1}m}\psi_{\gamma n j}$ (elements of the orbit with fixed
$\gamma$) we reproduce the above example of $p$--adic
multiresolution analysis.

In the general case $f\in D_0(\mathbb{Q}_p)$ the orbit by theorem
\ref{main_theorem} is given by the set
$\{G(p^{\gamma}J,p^{\gamma}Jn)f\}$.

Let us define the wavelet space $W_{-\gamma}$ as the linear span of
vectors $G(p^{\gamma}J,p^{\gamma}Jn)f$ with fixed $\gamma\in
\mathbb{Z}$. In this case the wavelet spaces $W_{\gamma}$,
$W_{\gamma'}$ are not necessarily orthogonal. The spaces
$W_{\gamma}$, $W_{\gamma'}$ are orthogonal if $|\gamma-\gamma'|$ is
sufficiently large, namely, it is larger than the difference of the
maximal and the minimal scales $\gamma_1$ of the wavelets in the
expansion
$$
f=\sum_{\gamma_1 n_1 j_1}C_{\gamma_1 n_1 j_1}\psi_{\gamma_1 n_1
j_1}.
$$
The property (\ref{prop_2}) for the spaces $W_{\gamma}$ will be
satisfied.

We see that the frame $\{G(p^{\gamma}J,p^{\gamma}Jn)f\}$ satisfies
the main properties of the $p$--adic multiresolution analysis and
satisfies some non--trivial generalization of the orthogonality
property. Therefore it is natural to consider the structure of this
frame as a generalization of the multiresolution construction.

In particular, in the case (\ref{fixed_scale}) where $f\in
D_0(\mathbb{Q}_p)$ is a linear combination of wavelets with a fixed
scale, the spaces $W_{\gamma}$ will be orthogonal.

\bigskip

\noindent{\bf Remark}\qquad The $p$--adic affine group can be
considered as the inverse (projective) limit of the factor groups
$$
G=\underleftarrow{\lim}\, G/G_f
$$
where the sequence of $f\in D_0(\mathbb{Q}_p)$ is such that the
corresponding stabilizers $G_f$ tend to the trivial subgroup in $G$
consisting of the unity element.

This limit corresponds to the joint limit $\gamma_A\to+\infty$,
$\gamma_0\to-\infty$, where $\gamma_A$, $\gamma_0$ are defined by
(\ref{stab_1}), (\ref{stab_2}) correspondingly.

The multiresolution construction corresponds to the element of the
sequence $\{G/G_f\}$ with $\gamma_A=1$, $\gamma_0=0$. In particular,
the corresponding $f$ can be chosen according to
$f(x)=\psi(x)=\chi(p^{-1}x)\Omega(|x|_p)$.

Therefore the multiresolution construction in $p$--adic analysis can
be naturally considered as a term in the sequence of the inverse
limit, and we can consider other terms in this sequence as
generalizations of the multiresolution analysis.

\section{Appendix}

Let us recall some constructions of $p$--adic analysis, see
\cite{VVZ} for details and, e.g. \cite{RandomWalk}, \cite{Andr} for
applications. The field $\mathbb{Q}_p$ of $p$--adic numbers is the
completion of $\mathbb{Q}$ with respect to the $p$--adic norm
$|\cdot|_p$, defined as follows. For any rational number we consider
the representation
$$
x=p^{\gamma}{m\over n}
$$
where $p$ is a prime, $\gamma$ is an integer, $p$, $m$, $n$ are
mutually prime numbers, $n\ne 0$. The $p$--adic norm is defined as
follows
$$
|x|_p=p^{-\gamma}.
$$
A $p$--adic number can be uniquely represented by the series
$$
x=\sum_{j=\gamma}^{\infty}x_jp^j,\qquad x_j=0,\dots,p-1
$$
which converges in the $p$--adic norm.

A complex valued character $\chi:\mathbb{Q}_p\to \mathbb{C}$ of a
$p$--adic argument (where $x$ has the form of the above series) is
defined by
$$
\chi(x)=\exp \left(2\pi i \sum_{j=\gamma}^{-1}x_j p^j\right),\qquad
\chi(x+y)=\chi(x)\chi(y).
$$
When the above $\gamma$ is nonnegative the sum above is equal to
zero and the character is equal to one.

The character $\chi$ is a locally constant function. The function
$f$ is locally constant if for any $x$ there exists $\varepsilon$
for which $\forall y$: $|x-y|_p\le \varepsilon$ we have $f(x)=f(y)$.

Another example of a locally constant function is the characteristic
function of a $p$--adic ball. We denote by $\Omega(x)$ the
characteristic function of the interval $[0,1]$.  The characteristic
function of the $p$--adic ball of radius 1 with center in 0 has the
form:
$$
\Omega(|x|_p)=\left\{
\begin{array}{rc}
1,&|x|_p\le 1\;,\\
0,&|x|_p>1\;.\\
\end{array}
\right.
$$
Let us note that, unlike in the real case, the characteristic
function of a $p$--adic ball is continuous (the same holds for an
arbitrary locally constant function).

The $p$--adic wavelet $\psi$ is the product of character and
characteristic function of the ball
$$
\psi(x)=\psi(p^{-1}x)\Omega(|x|_p).
$$

The Bruhat--Schwartz space $D(\mathbb{Q}_p)$ of $p$--adic test
functions is the linear space of locally constant complex valued
functions with compact support. Any function in $D(\mathbb{Q}_p)$ is
a (finite) linear combination of characteristic functions of balls.

The space $L^2(\mathbb{Q}_p)$ is the Hilbert space of complex valued
functions which are square integrable with respect to the Haar
measure on $\mathbb{Q}_p$. For the Haar measure on $\mathbb{Q}_p$
the volume of the ball is equal to the diameter of this ball.

The next theorem describes the basis of $p$--adic wavelets
\cite{wavelets}.

\begin{theorem}\qquad {\sl
The set of functions $\{\psi_{\gamma nj}\}$:
\begin{equation}\label{basispaw}
\psi_{\gamma nj}(x)=p^{-{\gamma\over 2}} \chi(p^{-1}j
(p^{\gamma}x-n)) \Omega(|p^{\gamma} x-n|_p),
\end{equation}
$$
\gamma\in \mathbb{Z},\quad j=1,\dots,p-1, \quad n\in
\mathbb{Q}_p/\mathbb{Z}_p,
$$
\begin{equation}\label{Qp/Zp}
n=\sum_{l=-\delta}^{-1}n_l p^{l},\quad n_l=0,\dots,p-1,
\end{equation}
is an orthonormal basis in $L^2(\mathbb{Q}_p)$ (the basis of
$p$--adic wavelets).}
\end{theorem}

In the present paper the notation $n\in \mathbb{Q}_p/\mathbb{Z}_p$
always means that we choose for $n$ the representative in the
corresponding equivalence class of the form (\ref{Qp/Zp}). More
generally, for $n\in \mathbb{Q}_p/p^{\gamma_0}\mathbb{Z}_p$ we
always choose the representative
$$
n=\sum_{l=-\delta}^{\gamma_0-1}n_l p^{l},\quad n_l=0,\dots,p-1.
$$

The space $D_0(\mathbb{Q}_p)$ is the space of mean zero test
functions. Any function in $D_0(\mathbb{Q}_p)$ is a (finite) linear
combination of $p$--adic wavelets:
$$
f=\sum_{\gamma n j}C_{\gamma n j}\psi_{\gamma n j},\qquad C_{\gamma
n j}\in \mathbb{C}.
$$

\begin{definition}\qquad {\sl
The set of vectors $\{f_n\}$ in the Hilbert space ${\cal H}$ is a
frame, if there exist positive constants $A,B>0$, such that for each
vector $g\in {\cal H}$ the following inequality is satisfied:
$$
A\|g\|^2\le\sum_{n} |\langle g,f_n\rangle|^2 \le B\|g\|^2.
$$
}
\end{definition}

The constants $A$ and $B$ are called the lower and the upper bounds
of the frame correspondingly. A frame is tight if the frame bounds
$A$ and $B$ are equal. A frame is uniform if all elements have equal
norms.

The unitary representation in $L^2(\mathbb{Q}_p)$ of the $p$--adic
affine group $G$ acts on the function $f\in L^2(\mathbb{Q}_p)$ by
translations and dilations
$$
G(a,b)f(x)=f^{a,b}(x)=|a|_p^{-{1\over 2}}f\left({x-b\over a}\right),
$$
where $a,b\in \mathbb{Q}_p$ and $a\ne 0$.

The composition of the elements of the affine group has the form
$$
G(a,b)G(a',b')=G(aa',b+ab').
$$
This implies for the degree of $G(a,b)$
$$
G^n(a,b)=G(a^n,b[n]_a).
$$
Here $[n]_1=n$ and for $a\ne 1$
$$
[n]_a={1-a^n\over 1-a}.
$$

The inverse to $G(a,b)$ has the form:
$$
G^{-1}(a,b)=G\left(a^{-1},-ba^{-1}\right).
$$

\bigskip

\noindent{\bf Acknowledgments}\qquad One of the authors (S.K.) would
like to thank I.V.Volovich, V.S.Vla\-di\-mi\-rov and A.Yu.Khrennikov
for fruitful discussions and valuable comments. He gratefully
acknowledges being partially supported by the grant DFG Project 436
RUS 113/809/0-1 and DFG Project 436 RUS 113/951, by the grants of
the Russian Foundation for Basic Research  RFFI 05-01-04002-NNIO-a
and RFFI 08-01-00727-a, by the grant of the President of Russian
Federation for the support of scientific schools NSh 3224.2008.1 and
by the Program of the Department of Mathematics of the Russian
Academy of Science ''Modern problems of theoretical mathematics''.
He is also grateful to IZKS (the Interdisciplinary Center for
Complex Systems) of the University of Bonn for kind hospitality.


\begin{thebibliography}{99}


\bibitem{Daubechies}  {\it I.Daubechies }
Ten Lectures on Wavelets, CBMS Lecture Notes Series. SIAM,
Philadelphia, 1992.

\bibitem{Meyer} {\it Y.Meyer} Wavelets and operators, Cambridge
University Press, Cambridge, 1992.

\bibitem{Mallat} {\it S.G. Mallat} A Wavelet Tour of Signal Processing, Academic Press,
1999

\bibitem{NPS} {\it I.Ya. Novikov, V.Yu. Protasov, M.A. Skopina} Wavelet Theory. Moscow:
Fizmatlit, 2005. (in Russian)

\bibitem{wavelets} {\it S.V.Kozyrev} Wavelet theory as $p$-adic spectral analysis   //
Izvestiya: Mathematics. 2002. V.66. N.2. P.367--376.
http://arxiv.org/abs/math-ph/0012019

\bibitem{Benedetto} {\it J.J. Benedetto,   R.L. Benedetto}  A wavelet theory
for local fields and related groups //  The Journal of Geometric
Analysis. 2004. V.14. N.3. P.423-456.

\bibitem{Benedetto1} {\it R.L.Benedetto} Examples of wavelets for local
fields // Contemporary Mathematics. 2004. V.345. AMS, Providence.
P.27-47. http://arxiv.org/math.CA/abs/0312038

\bibitem{ShelkSkop} {\it V.M.Shelkovich, M.Skopina}
$p$--Adic Haar multiresolution analysis,
http://arxiv.org/abs/0704.0736

\bibitem{KhrShelkSkop} {\it A.Yu. Khrennikov, V.M. Shelkovich, M. Skopina} $p$-Adic refinable functions
and MRA-based wavelets, http://arxiv.org/abs/0711.2820


\bibitem{AKhSh} {\it S.Albeverio, A.Yu.Khrennikov, V.M.Shelkovich}
Harmonic analysis in the $p$--adic Lizorkin spaces: fractional
operators, pseudo-differential equations, $p$--adic wavelets,
Tauberian theorems // The Journal of Fourier Analysis and
Applications. 2006. V.12. N.4. P.393-425.

\bibitem{cont_wavelets}
{\it S.Albeverio, S.V.Kozyrev} Coincidence of the continuous and
discrete $p$--adic wavelet transforms,
http://arxiv.org/abs/math-ph/0702010

\bibitem{VVZ}
{\it  V.S.Vladimirov,   I.V.Volovich,  Ye.I.Zelenov} $p$--Adic
Analysis and Mathematical Physics. Singapore: World Scientific,
1994.

\bibitem{RandomWalk} {\it S.Albeverio, W.Karwowski} A random walk on $p$--adics --- the
generator and its spectrum // Stoch. Processes Appl. 1994. V.53.
no.1. P.1--22.

\bibitem{Andr}  {\it A.Yu.Khrennikov, M.Nilsson} $P$--adic Deterministic and Random
Dynamics, Kluwer Academic, Dordrecht-Boston-London, 2004.













\end{thebibliography}
\end{document}